\documentclass[english,number,5p,sort&compress]{elsarticle}
\usepackage[T1]{fontenc}
\usepackage[latin9]{inputenc}
\usepackage{color}
\usepackage{amstext}
\usepackage{amssymb}
\usepackage{graphicx}

\makeatletter

\usepackage{babel}

\makeatother

\usepackage{babel}
\begin{document}

\title{Spin waves in planar quasicrystal of Penrose tiling}

\author[uam]{J. Rych\l y}

\author[uam]{S. Mieszczak}

\author[uam]{J.W. K\l os\corref{cor1} }

\ead{klos@amu.edu.pl}

\cortext[cor1]{Corresponding author}

\address[uam]{Faculty of Physics, Adam Mickiewicz University in Poznan, Umultowska
85, Pozna\'{n}, 61-614, Poland}

\begin{frontmatter}{} 
\begin{abstract}
We investigated two-dimensional magnonic structures which are the
counterparts of photonic quasicrystals forming Penrose tiling. We
considered the slab composed of Ni (or Py) disks embedded in Fe (or
Co) matrix. The disks were arranged in quasiperiodic Pernose-like
structure. The infinite quasicrystal was approximated by its rectangular
section with periodic boundary conditions applied. This approach allowed
us to use the plane wave method to find the frequency spectrum of
eigenmodes for spin waves and their spatial profiles. The calculated
integrated density of states shows more distictive magnonic gaps for
the structure composed of materials of high magnetic contrast (Ni
and Fe) and relatively high filling fraction. This proves the impact
of quasiperiodic long-range order on the spectrum of spin waves. We
also investigated the localization of SW eingenmodes resulting from
the quasipeiodicity of the structure. 
\end{abstract}
\begin{keyword}
spin waves, magnonics, quasicrystals, Penrose tiling 
\PACS 75.30.Ds, 75.50.Kj, 75.75.-c, 75.78.-n 
\end{keyword}
\end{frontmatter}{}

\section{Introduction}

The spectrum of wave excitations reflects the structural properties
of the system. It is known, that the long-range order in periodic
and quasiperiodic structures can be revealed by the presence of forbidden
frequency gaps in the spectrum \citep{Steu07,Janot94}. The quasiperiodic
system have more complex band structure resulting from countable set
of a Bragg peaks densely filling reciprocal space \citep{Stein87,Var13}
and, connected to them, frequency gaps. Due to this feature, the spectrum
of scattered waves from quasiperiodic structures can have fractal
structure \citep{Rych15,Zhu05,Gell94} which can be used for advanced
signal filtering and processing \citep{Mac12}.

In quasiperiodic system, there is a possibility to obtain omnidirectional
frequency gap by optimizing the structure with rotational symmetry,
which is unique property of quasicrystals \citep{Bay01a,Jia11,Rech08,Sutter07}.
It is fundamentally different for periodic structures of the same
contrast of constituent materials which can be also useful for application.

The quasiperiodic structures are also interesting because of different
localization mechanisms than in periodic systems. The Bloch waves
in periodic structures are spatially extended in the absence of defects
and surfaces, whereas the eigenmodes in quasiperiodic system can be
localized \citep{Rych15,Gell94} in the bulk region of the structure.
For the system with translational symmetry every unit cell is equivalent
and there is no reason for localization. For self-similar quasicrystals,
the system form the hierarchical structure in which the localization
can be expected.

The rich spectrum of the gaps and the increase of localization can
lead to the strong suppression of group velocity \citep{Neg03}. This
affects (deteriorates) the transport properties in quasi- periodic
structures. Surprisingly, the introduction into quasi- periodic structure,
a particular amount of structural defects (Anderson localization regime)
can cause the disorder- enhanced transport \citep{Lev11}.

The dynamical properties of quasiperiodic composite structures were
investigated for different kind of media \citep{Alb03,Steu07} \textendash{}
electronic \citep{Ma89,Mer85}, photonic \citep{Kali00,Var13,Koh87},
plasmonic \citep{Nam16}, phononic \citep{Chen08,Sutter07} and magnonic
systems \citep{Cost13,Liu93,Bhat14}. For alll kinds of mentioned
media, the two-dimensional (2D) quasiperiodic structures \citep{Kali00,Vig16,Bay01b,Chan98,Ma89}
give more possibility to adjust structural parameters and to mold
the spectrum of excitation than the one-dimensional (1D) quasiperiodic
structures. Therefore, we focused our studies on spin wave (SW) dynamics
in 2D magnonic quasicrystals, which is promising but not extensively
explored subject.

The most commonly considered 2D quasiperiodic structure is Penrose
tiling \citep{Steu09}. SW dynamics in magnonic quasicrystals in the
form of Penrose-like structure were, up to know, considered mostly
for lattice models, in which a Heisenberg antiferromagnet model was
investigated \citep{Vedm03,Jaga12}. This system, for inhomogeneous
Néel-ordered ground state, shows the presence of frequency gaps in
the frequency spectrum of SWs \citep{Szall08,Szall09}. The studies
of SW excitation in the 2D magnonic quasiperiodic structures, in which
the richness of structural and material factors play significant role,
are on initial stage.

The subject of SW dynamics in quasiperiodically patterned structures
is almost unexplored. However, some interesting reports on magnetic
antidotes (quasi- periodic) lattices, forming the P2 Penrose coverage
\citep{Bhat13} or Ammann tailing \citep{Bhat14} were published in
the last few years. The structures considered in the studies \citep{Bhat13,Bhat14}
have large filling fraction (the small volume fraction of magnetic
material). They have form of network of magnetic wires of sizes in
the range of single micrometers for one section of the network. In
this crossover dipolar-exchange regime, the shape anisotropy results
in significant static and dynamic magnetic fields \citep{Klos2012,Tac15}.
The demagnetizing effects influence significantly both magnetic configuration
and SW dynamics. Strongly anisotropic SW dependence on the direction
of magnetic field is observed even for in-phase precession of SWs
(investigated by ferromagnetic resonance (FMR)). This anisotropy is
absent in an exchange regime, in which demagnetizing fields are negligible.
For the studies on magnonic quasicrystals presented in \citep{Bhat13,Bhat14},
it is challenging to deduce what is the impact of quasiperiodicity
on the SWs dynamics. The reported anisotropy can be related both to
the quasiperiodic ordering (resulting in Bloch scattering) and to
the shape of large antidots.

The clear signature of coherent wave dynamics on (quasi)- periodicity
is Bragg scattering, which can lead to the opening of the frequency
gaps. Our aim is to investigate the impact of quasiperiodicity on
the spectral characteristics and localization properties of SWs in
2D structures.

We would like to minimize the shape anisotropy effects. Therefore,
we will consider the system in which the long-range order can be manifested
by the presence of magnonic gaps. In these systems the SW scatering
on quasiperiodic lattice can lead to the enhancement of localization.
To observe frequency gaps in SW spectrum of 2D quasicrystal nanosucture
(in the form of Penrose structure), we considered the bi-component
planar magnonic quasicrystals, in which the exchange coupling between
magnetic nanoelements is mediated by the matrix. The strong coupling
between inclusions and high contrast of magnetic parameters between
inclusions and matrix is beneficial for strong SW scattering. It also
makes the observation of frequency gaps more feasible. The isotropic
shape of inclusions (disks), their small sizes and distances between
them, make the exchange interactions overshadowing the dipolar ones
and reduce the static shape anisotropy effects.

\section{Structure and model}

\begin{figure}[!ht]
\centering{}\includegraphics[width=0.75\columnwidth]{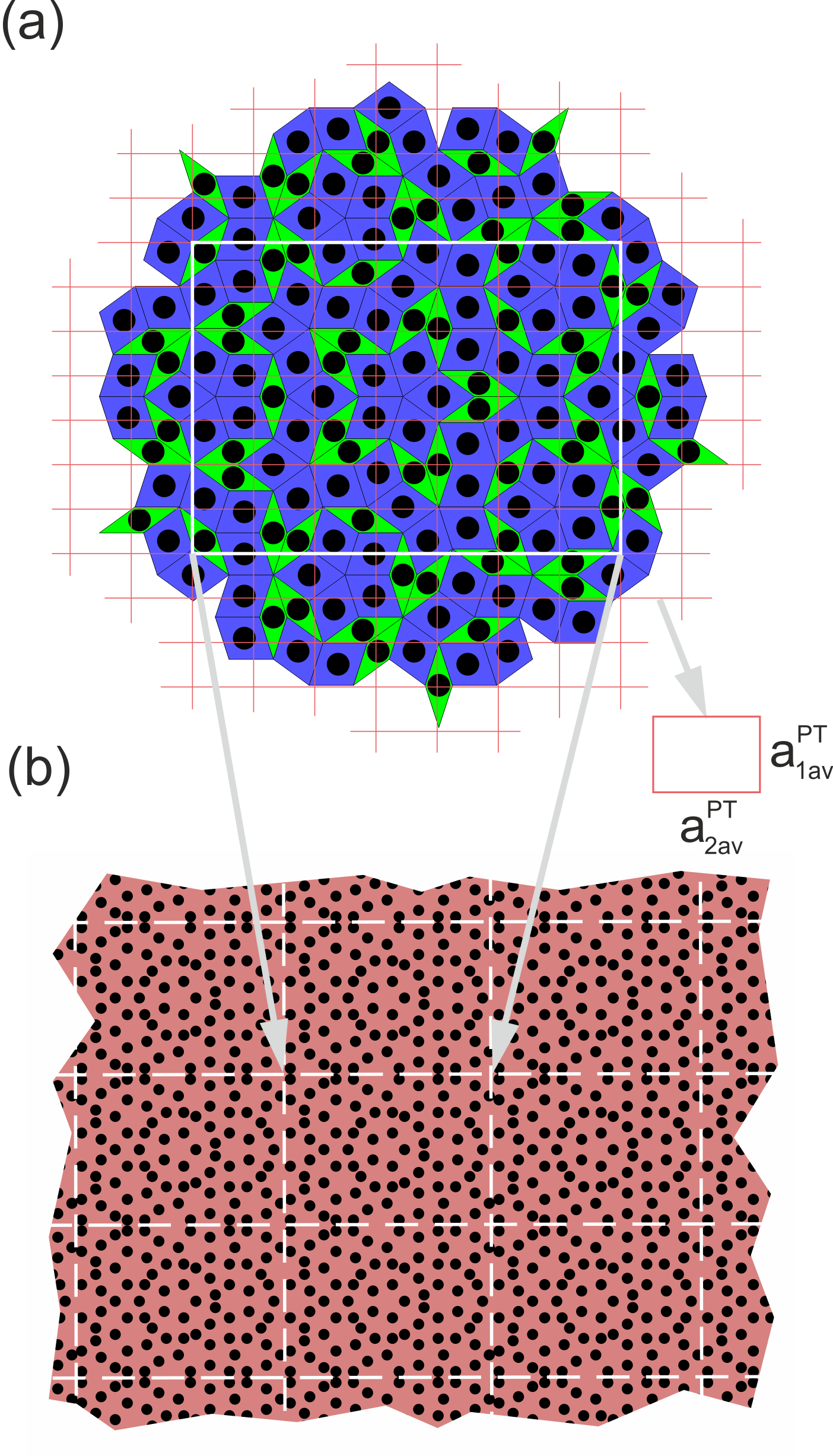} \caption{(a), (b) The Penrose P3 tiling in the form of rosette of 5-fold symmetry.
The considered magnonic structure consists of Ni (or Py) disks of
diameter $5.6$ nm embedded in Fe (or Co) slab of the same thickness:
$2\text{ nm}$. The disks are placed in the centers of Penrose tiles:
wide (blue) and narrow (green) rhombus of the side $\sim10.85\text{ nm}$.
The orthorhombic cells, marked by the sets of red lines, present the
regular grids of periods equal to averaged periods of Penrose tiling
\citep{Steu99} with two vertices per unit cell in average. The centered
orthorhombic unit cells were used to build rectangular supercell for
the plane wave method (PWM) calculations}
\end{figure}

We investigated propagation of SWs in 2D bicomponent magnonic structure
in the form of Penrose-like quasicrystal. The structure is based on
P3 Penrose coverage \citep{Kali00}, constructed from two rhombi tiles.
Every rhombus has the same lengths of sides, but different acute angle:
$\alpha=\pi/5$ for narrower, $\alpha=2\pi/5$ for wider one. For
such definition, the ratio between areas of the mentioned two rhombi
is $1:\tau$, i.e. golden ratio. In the center of every rhombus, the
inclusions in a form of a disks are placed \citep{Kali00,Not2004}.
These ferromagnetic disks are embedded in the plane of the same thickness
made of the different kind of ferromagnetic material. The lengths
of the sides of rhombi are $\sim10.85\text{ nm}$ and the radii of
disks are $5.6\text{ nm}$, which give us filling fraction equal to
$f\!f=0.258$ (calculated for the largest supercell). For magnonic
crystals in an exchange dominated regime \citep{Klos2012}, the widest
frequency gaps in the SW spectrum occur for $f\!f\sim0.5$. To approach
this range, we assumed the filling fraction close to the maximal possible
value for considered structure, for which the inclusions do not overlap
with each other (see Fig. 1(a)). We premised the thickness equal to
$2\text{ nm}.$ For the small value of the ratio between thickness
and in-plane dimensions of the structure (diameter of inclusions and
distances between them), we can treat the system as two-dimensional
by avoiding the quantization. To minimize the static dipolar effects,
related to the shape anisotropy and to investigate the system in an
exchange dominated regime, we assumed relatively small dimensions
of the elements composing the structure \citep{Kraw2012,Klos2012}.
The gyromagnetic ratio equals $\gamma=176\text{ GHz/T}$ in both materials.
For matrix we selected two materials, Fe or Co characterised by a
saturation magnetization $M_{\rm{S}}$ and an exchange length $\lambda_{\rm{ex}}$,
that equal to $M_{\rm{S,Fe}}=1.752\cdot10^{6}\text{ A/m}$, $\lambda_{\rm{Fe}}=3.30\text{ nm}$,
$M_{\rm{s,Co}}=1.445\cdot10^{6}\text{ A/m}$, $\lambda_{\rm{Co}}=4.78\text{ nm}$,
accordingly. For inclusions we took Ni or Py where: $M_{\rm{S,Ni}}=0.484\cdot10^{6}\text{ A/m}$,
$\lambda_{\rm{Ni}}=7.64\text{ nm}$, $M_{\rm{S,Py}}=0.860\cdot10^{6}\text{ A/m}$,
$\lambda_{\rm{Py}}=5.29\text{ nm}$. We combined two pairs of materials:
Fe/Ni and Co/Py. We assumed that our sample is saturated by an external
field, which value equals to $\mu_{0}H_{0}=0.2\text{ T}$.

The precession of the magnetization vector is described by Landau-Lifshitz
equation (LLE):

\begin{equation}
\frac{\partial\mathbf{M}}{\partial t}=\mu_{0}\gamma\mathbf{M}\times\mathbf{H}_{\rm{\rm{ef\!f}}},
\end{equation}
where $\mu_{0}$ is the permeability of vacuum, $\mathbf{M}_{\rm{S}}$
is saturation magnetization. The effective magnetic field $\mathbf{H}_{\rm{ef\!f}}$
is composed of the following terms:

\begin{equation}
\mathbf{H}_{\rm{ef\!f}}\left(\mathbf{r},t\right)=\mathbf{H}_{0}+\mathbf{H}_{\rm{dm}}\left(\mathbf{r},t\right)+\mathbf{H}_{\rm{ex}}\left(\mathbf{r},t\right),
\end{equation}
where $\mathbf{H}_{0}$ means external field, $\mathbf{H}_{\rm{dm}}\left(\mathbf{r},t\right)$
is demagnetizing field and $\mathbf{H}_{\rm{ex}}\left(\mathbf{r},t\right)$
is exchange field. We have solved (LLE) using plane wave method (PWM)
\citep{Rych15b}.

\begin{figure*}[!ht]
\centering{}\includegraphics[width=1\textwidth]{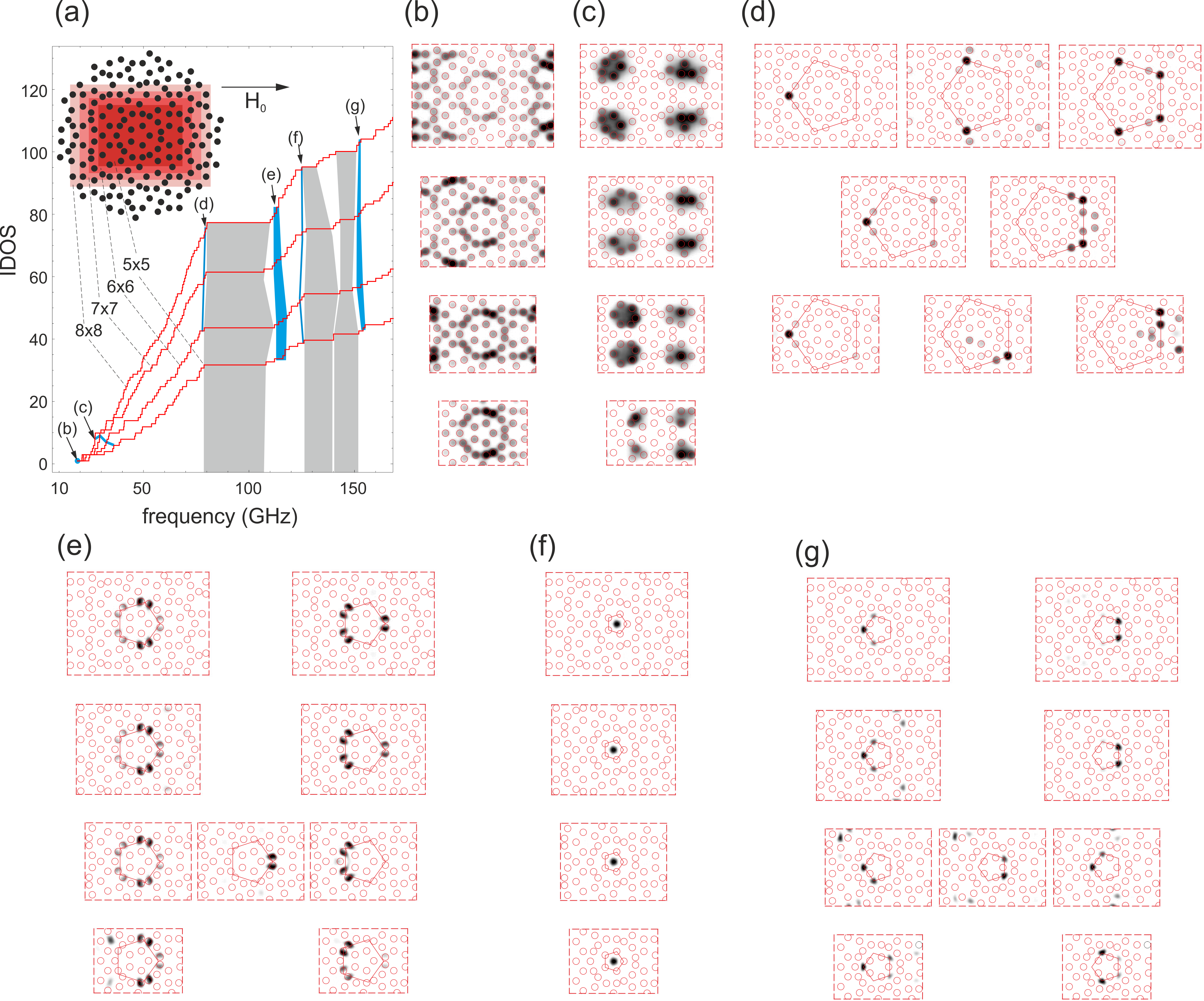} \caption{(a) Integrated density of states calculated for SWs in magnonic Penrose-like
structure (approximated by supercells of different sizes \textendash{}
composed of 5x5, 6x6, 7x7, and 8x8 orthorhombic cells of averaged
periods - see the inset). Gray areas mark the widest magnonic gaps.
The magnetic field (0.2 T) was applied along the longer side of supercells.
The spatial distribution of the amplitude of out-of-plane component
of dynamical magnetization are presented in (b)-(g) for selected eigenmodes. }
\end{figure*}

The PWM requires periodic structures. Therefore, we use the large
supercells with periodic boundary conditions as approximates of the
Penrose-like structure. To reduce the impact of artificial structural
defect on the edges of the supercells, we constructed the supercells
as rectangular arrays of centered orthorhombic cells (see Fig. 1)
of the sizes being few averaged periods of the Penrose lattice \citep{Steu99}
with:

\begin{equation}
a_{\rm{1\,av}}^{\rm{PT}}=\left(3-\tau\right)a_{\rm{r}},\ \ a_{\rm{2\,av}}^{\rm{PT}}=\left(3-\tau\right)^{3/2}a_{\rm{r}}.
\end{equation}
where $a_{\rm{r}}$ is the edge length of the rhombus tile (see Fig. 1(a)).

\section{Results and discussion }

\begin{figure*}[ht]
\centering{}\includegraphics[width=1\textwidth]{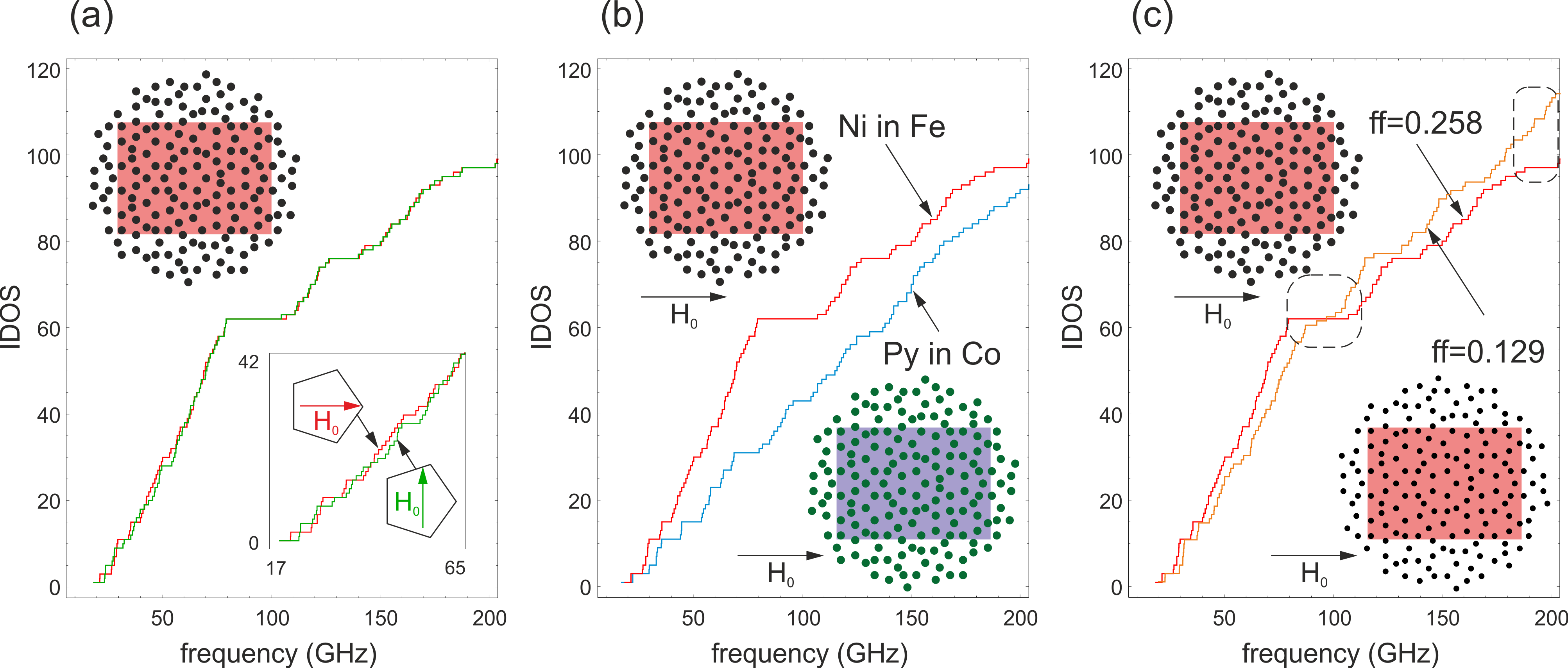} \caption{Impact of: (a) direction of the magnetic field, (b) contrast of the
magnetic materials and (c) filling fraction on the SW spectrum in
magnonic Penrose-like structure. The change of the direction of magnetic
field between two nonequivalent directions affects only slightly the
lowest SW modes (we kept the magnitude of the field constant: 0.2
T which is strong enough to saturate the sample). Magnonic gaps in
the structure composed of materials with lower magnetic contrast (Py
inclusions in Co matrix) are narrower (b). The similar effect - decreasing
the size of some magnonic gaps (see the regions marked by dashed boxes
in (c) ), can be observed also for the structures with the lower filling
fractions (c).}
\end{figure*}

We have approximated the Penrose tiling by the periodic array of rectangular
supercells, containing 5x5, 6x6, 7x7, 8x8 small orthorhombic cells
\citep{Steu99} for each supercell. Each orthorhombic cell has the
sizes being the effective periods of Penrose lattice in two orthogonal
directions. From the calculations, we have obtained the eigenvalues
(frequencies of SWs) and the corresponding eigenvectors (profiles
of SW modes). We have ordered obtained SW modes by increasing eigenfrequencies,
and then we calculated the integrated density of states (IDOS) \citep{Rych15}.
IDOS spectra (IDOS as a function of frequency), calculated for the
supercells of different sizes, are shown in Fig. 2(a). In IDOS spectra
we have found a few distinctive frequency gaps. Magnonic frequency
gaps can be identified by finding the plateau regions in the IDOS
spectrum. For these frequencies, bulk modes do not occur and because
of that, IDOS is not increasing (these gaps are marked by grey areas
in Fig. 2(a)). Clear identification of frequency gaps in the same
frequency ranges for the successive approximates of the Penrose structure
(supercells) is an evidence of that the approximates are large enough
to reveal the quasiperiodic long-range order. For consecutive approximations,
in the same ranges of frequencies, we have also found similar SW mode
profiles. 

In the Fig. 2(b-g) are presented profiles of SW modes for the system
with the supercells of successive sizes. They are grouped accordingly
to similar shapes of their spatial distributions. The frequencies
of selected modes are marked on the SW spectra (Fig. 2(a)) by blue
stripes. The modes of the lowest frequencies have to be discussed
separately. By virtue of the fact that on the boundaries of the unit
cells the periodic boundary conditions were used, the modes of the
lowest frequencies are quantized in the area of the whole supercell.
In the Fig. 2(b) are presented profiles of the SW modes of the lowest
frequency in the supercells of different sizes. This mode is concentrated
mainly in the Ni inclusions and do not have any nodal lines (each
Ni inclusion is excited more or less with the same strength). Among
the lowest modes we selected also the mode with one nodal line in
each of the two perpendicular directions (Fig. 2(c)). The frequency
of these modes are dependent on the size of the supercell. It proves
that by using the PWM in a supercell approach we cannot rigorously
investigate low frequency excitations, which spatial distribution
is comparable to the sizes of supercells.

For higher frequnecies, the modes of corresponding spatial profiles
have very similar frequencies for successive approximates of the Penrose
structure. Their amplitudes start to localize at the particular regions
of the structure. The largger the supercell is, the better the five-fold
symmetry is restored and the lesser is the impact of the rectangular
shape of the supercell on the SW spectra. In the Fig. 2(d) are presented
modes found beneath the first big band gap. Those modes are concentrated
in the strictly chosen inclusions of Ni disks, which which are characterized
by the lower FMR frequency then Fe matrix. We can find two or three
modes of that kind, which appear at the vertices of the same regular
pentagon. For the small supercells, modes are concentrated at the
individual/single Ni disks, whereas for the larger supercells we can
find the state which concentrates at the four of five available \textcolor{black}{apices}
of the pentagon. It can be easily deduce that by making bigger Penrose
structure and by choosing bigger orthorhombic cell of averaged period,
it would be possible to find a mode that is concentrated exactly at
all five \textcolor{black}{apices} of a mentioned pentagon (at the
Ni disks). The next group of the profiles, presented in the Fig. 2(e),
refer to the modes of the frequencies just above the biggest frequency
gap. In these profiles, the SWs amplitude is localized in few pairs
of spots distributed in the vertices of pentagon. Each pair occupy
the two, closest to each other, disks (located in the centers of the
pairs of narrower rhombi - see Fig. 1(a)). This pentagon is smaller
one than the pentagon from Fig. 2(d). Modes from the Fig. 2(f) and
Fig. 2(g) appear below the second and above the third, recognized
by us, band gap, accordingly.These modes have the frequencies above
the FMR of Fe. Therefore, they start to concentrate in the material
of Fe matrix, in the void areas surrounded by the Ni disks. The modes
from Fig. 2(f) are concentrated exactly in the middle of the smallest
pentagon - i.e. in the area surrounded by the five Ni disks placed
in the centers of the wider rhombuses. Due to the size of the supercell,
we have only single location of this kind. The modes appearing above
the third big band gap are concentrated also in the template material
between the Ni disks. These modes are concentrated in such a way that
they also form a pentagon, surrounding the middle of the structure,
the center of the main pentagon, in which the modes from Fig. 2(f)
are concentrated.It is worth to notice that all modes appear in the
vertices of pentagons, which center is located exactly at the point,
in which mode from the Fig. 2(f) is concentrated.\textbf{ }The modes
from Fig. 2(g) create the small pentagon, surrounding discussed point.
However, the \textcolor{black}{apices} of those modes are not concentrated
on the Ni disks, but between them, the Ni disks are placed in the
middle of the edges of this pentagon.

We investigated the nanostructure in the form of the quasicrystal,
composed of matrix with small inclusions placed close to each other.
We expect that the static demagnetizing field, dependent on the direction
of the applied magnetic field with respect to the structure, will
be negligible inan exchange regime. To check this prediction, we calculated
IDOS for the two different directions of magnetic field (with respect
to the structure), which is presented in Fig. 3(a). We observe only
the tiny differences of IDOS in the low frequency range. Positions
of the largest frequency gaps and the higher parts of the spectra
are practically unaffected by the change of the direction of the magnetic
field. It means that the change of the direction of an applied field
do not influences SW dynamics. The independence of IDOS on the direction
of external magnetic field do not mean that the SWs propagation is
isotropic in quasiperiodic structures. Note that, for fixed magnetic
configuration, IDOS collects the contributions of all states for different
possible directions of propagation. 

In the SW spectrum presented in Fig. 2(a), we can identify a few wide
frequency gaps. However, we deliberately selected the values of structural
and material parameters which support the opening of frequency gaps.
If we choose the constituent materials of lower contrast of magnetic
parameters, then the frequency gaps are shrinked and can be too narrow
to be clearly identified (see, the IDOS spectrum for Py inclusions
in Co matrix in Fig. 3(b)). We chose also the structure of quite large
filling fraction $f\!f$ of Ni inclusions. It is understandable that
for the limiting cases, $f\!f=0$ and $f\!f=1$, we deal with uniform
material in which the frequency gaps are not observed. Therefore,
the intermediate value of filling fraction is optimal to obtain the
widest frequency gaps. For the selected value $f\!f=0.258$, the disks
are in close proximity. This geometrical constraint makes the further
increase of the filling fraction practically impossible. The widths
of the gaps, which we obtained for this value of filling fraction,
seem to be maximal. We can notice (see Fig. 3(c)) that the decrease
of filling fraction bellow $f\!f=0.258$ reduces the size of frequency
gaps.

\section{Conclusions }

We have calculated IDOS spectrum for SWs in Penrose-like planar bi-componet
magonic quasicrystal using PWM, in a supercell approach. We used the
supercell of the size being the multiplicity of the averaged period
of Penrose tiling. This reduces the strength of the defects introduced
by the boundaries of supercells, resulting from the presence of void
spaces or overlapped inclusions (see Fig. 1(b)) and make the supercell
method suitable to get an approximate solutions of quasicrystal.

In the calculated spectrum, we identified frequency gaps, which are
visible as plateaus in IDOS spectrum, appearing in the same frequency
ranges for the different sizes of supercells. The presence of frequency
gaps in the SW spectrum reveales the long range order in the quasicrystal
structure. However, the size of the supercell affects the spectrum
of modes of the lowest frequencies. For these modes, the spatial changes
of the amplitude are much larger than distances between neighboring
inclusions (in a quasicrystal structure) and are comparable to the
size of a supercell \textendash{} see Fig. 2(c). The frequency of
the mode of a particular symmetry, with respect to the whole supercell,
depends on the size of supercell in this regime - see the blue lines
in Fig. 2(a), marking the frequencies of the modes with one horizontal
and one vertical line, which are presented in Fig. 2(c). The profiles
of the higher-frequency modes become more localized and start to reflect
structure of the quasicrystal. Their frequencies are converged for
larger supercells. However, the profiles of these modes barely preserve
a fivefold symmetry, due to rectangular shape of the supercell. They
are localized, with increasing frequency, in: the selected Ni inclusions
(the material with lower FMR frequency) (Fig. 2(d)), in the doublets
of Ni inclusions (Fig. 2(e)) and within the void spaces of Fe matrix
(Fig. 2(g)). For the considered sizes of the structure (diameter of
inclusions, thickness), the modes of higher frequencies are in an
exchange regime. Their frequencies are quite robust on the changes
of the direction of external field. The small anisotropy is observed
for the lowest frequency modes (see Fig. 3(a)). The magnonic gaps,
which we found, are one of the signatures of the long-range order
in the considered planar magnonic quasicrystals. We were able to identify
them clearly for the structures with relatively high contrast of magnetic
properties of constituent materials and for quite large value of filling
fraction, in which the inclusions are in the close proximity (see
Fig. 3(b,c)).

\section*{Acknowledgement}

The authors would like to thank Prof. M. Krawczyk for fruitful discussion
and comments.

This project has received funding from the European Union's Horizon
2020 research and innovation program under Marie Sk\l odowska-Curie
Grant Agreement No. 644348 and from the Polish National Science Centre,
project UMO-2012/07/E/ST3/00538.

\section*{References}

 \bibliographystyle{elsarticle-num}
\bibliography{bib}

\begin{thebibliography}{10}
\expandafter\ifx\csname url\endcsname\relax
  \def\url#1{\texttt{#1}}\fi
\expandafter\ifx\csname urlprefix\endcsname\relax\def\urlprefix{URL }\fi
\expandafter\ifx\csname href\endcsname\relax
  \def\href#1#2{#2} \def\path#1{#1}\fi

\bibitem{Steu07}
{W. Steurer}, {D. Sutter-Widmer}, Photonic and phononic quasicrystals, Phys. D:
  Appl. Phys. 40 (2007) R229.

\bibitem{Janot94}
{C. Janot}, Quasicrystals, New York: Oxford University Press, 1994.

\bibitem{Stein87}
{P. J. Steinhardt}, { S. Ostlund}, Quasicrystals, Singapure: World Scientific
  Publishings, 1987.

\bibitem{Var13}
{Z. V. Vardeny}, {A. Nahata}, {A. Agrawal}, Optics of photonic quasicrystals,
  Nature Photonics 7 (2013) 177.

\bibitem{Rych15}
{J. Rych\l{}y}, {J. W. K\l{}os}, {M. Mruczkiewicz}, {M. Krawczyk}, Spin waves
  in one-dimensional bicomponent magnonic quasicrystals, Phys. Rev. B 92 (2015)
  054414.

\bibitem{Zhu05}
{S.V. Zhukovsky}, {A.V. Lavrinenko}, Spectral self-similarity in fractal
  one-dimensional photonic structures, Photonics and Nanostructures -
  Fundamentals and Applications 3 (2005) 129.

\bibitem{Gell94}
{W. Gellermann}, {M. Kohmoto}, {B. Sutherland}, {P. C. Taylor}, Localization of
  {L}ight {W}aves in {F}ibonacci {D}ielectric {M}ultilayers, Phys. Rev. Lett.
  72 (1994) 633.

\bibitem{Mac12}
{E. Maci\'{}a}, Exploiting aperiodic designs in nanophotonic devices, Rep.
  Prog. Phys 75 (2012) 036502.

\bibitem{Bay01a}
{M. Bayindir}, {E. Cubukcu}, {I. Bulu}, {E. Ozbay}, Photonic band gaps and
  localization in two-dimensional metallic quasicrystals, Europhys. Lett. 56
  (2001) 41.

\bibitem{Jia11}
{L. Jia}, {I. Bita}, {E. L. Thomas}, Photonic density of states of
  two-dimensional quasicrystalline photonic structures, Phys. Rev. A 84 (2011)
  023831.

\bibitem{Rech08}
{M. C. Rechtsman}, {H.-Ch. Jeong}, {P. M. Chaikin}, {S. Torquato}, {P. J.
  Steinhardt}, Optimized {S}tructures for {P}hotonic {Q}uasicrystals, Phys.
  Rev. Lett. 101 (2008) 07390.

\bibitem{Sutter07}
{D. Sutter-Widmer}, {W. Steurer}, Prediction of band gaps in phononic
  quasicrystals based on single-rod resonances, Phys. Rev. B 75 (2007) 134303.

\bibitem{Neg03}
{L. Dal Negro}, {C. J. Oton}, {Z. Gaburro}, {L. Pavesi}, {P. Johnson}, {A.
  Lagendijk}, {R. Righini}, {M. Colocci}, {D. S. Wiersma}, Light {T}ransport
  through the {B}and-{E}dge {S}tates of {F}ibonacci {Q}uasicrystals, Phys. Rev.
  Lett. 90 (2003) 05501.

\bibitem{Lev11}
{L. Levi}, {M. Rechtsman}, {B. Freedman}, {T. Schwartz}, {O. Manela}, {M.
  Segev}, Disorder-{E}nhanced {T}ransport in {P}hotonic {Q}uasicrystals,
  Science 332 (2011) 1541.

\bibitem{Alb03}
{E. L. Albuquerque}, {M. G. Cottam}, Theory of elementary excitations in
  quasiperiodic structures, Phys. Rep. 376 (2003) 225.

\bibitem{Ma89}
{P. Ma}, {Y. Liu}, Structure of the energy spectrum for a two-dimensional
  quasicrystal: {P}erturbation method, Phys. Rev. B 39 (1989) 10658.

\bibitem{Mer85}
{R. Merlin}, {K. Bajema}, {Roy Clarke}, {F. -Y. Juang}, {P. K. Bhattacharya},
  Quasiperiodic {G}a{A}s-{A}l{A}s {H}eterostructures, Phys. Rev. Lett. 55
  (1985) 1768.

\bibitem{Kali00}
{M. A. Kaliteevski}, {S. Brand}, {R. A. Abram}, {T. F. Krauss}, {R. DeLa Rue},
  {P. Millar}, Two-dimensional {P}enrose-tiled photonic quasicrystals: from
  diffraction pattern to band structure, Nanotechnology 11 (2000) 274.

\bibitem{Koh87}
{M. Kohmoto}, {B. Sutherland}, {K. Iguchi}, Localization of optics:
  {Q}uasiperiodic media, Phys. Rev. Lett. 58 (1987) 2436.

\bibitem{Nam16}
{F. A. Namin}, {Yu. A. Yuwen}, {L. Liu}, {A. H. Panaretos}, {D. H. Werner}, {T.
  S. Mayer}, Efficient design, accurate fabrication and effective
  characterization of plasmonic quasicrystalline arrays of nano-spherical
  particles, Sci. Rep. 6 (2016) 22009.

\bibitem{Chen08}
{A. -Li. Chen}, {Y.-S. Wang}, {Y.-F. Guo}, {Z.-D. Wang}, Band structures of
  {F}ibonacci phononic quasicrystals, Solid State Commun. 145 (2008) 103.

\bibitem{Cost13}
{C. H. O. Costa}, {M. S. Vasconcelos}, Band gaps and transmission spectra in
  generalized fibonacci $\sigma$(p,q) one-dimensional magnonic quasicrystals,
  Phys.: Condens. Matter 25 (2013) 286002.

\bibitem{Liu93}
{T. S. Liu}, {G. Z. Wei}, Spin waves in quasiperiodic layered ferromagnets,
  Phys. Rev. B 48 (1993) 7154.

\bibitem{Bhat14}
{V. S. Bhat}, {J. Sklenar}, {B. Farmer}, {B. Farmer}, {J. B. Ketterson}, {J. T.
  Hastings}, {L. E. De Long}, Ferromagnetic resonance study of eightfold
  artificial ferromagnetic quasicrystals, J. Appl. Phys. 115 (2014) 17C502.

\bibitem{Vig16}
{P. Vignolo}, {M. Bellec}, {J. B\"ohm}, {A. Camara}, {J-M. Gambaudo}, {U.
  Kuhl}, {F. Mortessagn}, Energy landscape in a {P}enrose tiling, Phys. Rev. B
  93 (2016) 075141.

\bibitem{Bay01b}
{M. Bayindir}, {E. Cubukcu}, {I. Bulu}, {E. Ozbay}, Photonic band-gap effect,
  localization, and waveguiding in the two-dimensional {P}enrose lattice, Phys.
  Rev. B 63 (2001) 161104(R).

\bibitem{Chan98}
{Y. S. Chan}, {C. T. Chan}, {Z. Y. Liu}, Photonic {B}and {G}aps in {T}wo
  {D}imensional {P}hotonic {Q}uasicrystals, Phys. Rev. Lett 80 (1998) 956.

\bibitem{Steu09}
{Walter Steuer}, {Sofia Deloudi}, Crystallography of quasicrystals. {C}oncepts,
  methods and structures, Springer, 2009.

\bibitem{Vedm03}
{E. Y. Vedmedenko}, {H. P. Oepen}, {J. Kirschner}, Decagonal
  {Q}uasiferromagnetic {M}icrostructure on the {P}enrose {T}iling, Phys. Rev.
  Lett. 90 (2003) 137203.

\bibitem{Jaga12}
{A. Jagannathan}, Quasiperiodic {H}eisenberg antiferromagnets in two
  dimensions, Eur. Phys. J. B 85 (2012) 68.

\bibitem{Szall08}
{A. Szallas}, {A. Jagannathan}, Spin waves and local magnetizations on the
  {P}enrose tiling, Phys. Rev. B 77 (2008) 104427.

\bibitem{Szall09}
{A. Szallas}, {A. Jagannathan}, {S. Wessel}, Phason-disordered two-dimensional
  quantum antiferromagnets, Phys. Rev. B 79 (2009) 172406.

\bibitem{Bhat13}
{V. S. Bhat}, {J. Sklenar}, {B. Farmer}, {J. Woods}, {J. T. Hastings}, {S. J.
  Lee}, {J. B. Ketterson}, {L. E. De Long}, Controlled {M}agnetic {R}eversal in
  {P}ermalloy {F}ilms {P}atterned into {A}rtificial {Q}uasicrystals, Phys. Rev.
  Lett 111 (2013) 077201.

\bibitem{Klos2012}
{J.W. K\l{}os}, {M.L. Sokolovskyy}, {S. Mamica}, {M. Krawczyk}, The impact of
  the lattice symmetry and the inclusion shape on the spectrum of 2{D} magnonic
  crystals, J. Appl. Phys. 111 (2012) 123910.

\bibitem{Tac15}
{S. Tacchi}, {P. Gruszecki}, {M. Madami}, {G. Carlotti}, {J. W. K\l{}os}, {M.
  Krawczyk}, {A. Adeyeye}, {G. Gubbiotti}, Universal dependence of the spin
  wave band structure on the geometrical characteristics of two-dimensional
  magnonic crystals, Sci. Rep. 5 (2015) 1036.

\bibitem{Steu99}
{W. Steurer}, {T. Haibach}, The periodic average structure of particular
  quasicrystals, Acta Crystallogr. A 55 (1998) 48.

\bibitem{Not2004}
M.~Notomi, H.~Suzuki, T.~Tamamura, K.~Edagawa, Lasing action due to the
  two-dimensional quasiperiodicity of photonic quasicrystals with a penrose
  lattice, Phys. Rev. Lett. 92 (2004) 123906.

\bibitem{Kraw2012}
{M. Krawczyk}, {M.L. Sokolovskyy}, {J.W. K\l{}os}, {S. Mamica}, On the
  formulation of the exchange field in the {L}andau-{L}ifshitz equation for
  spin-wave calculation in magnonic crystal, Adv. Cond. Matter Phys. 2012
  (2012) 764783.

\bibitem{Rych15b}
{J. Rych\l{}y}, {P. Gruszecki}, {M. Mruczkiewicz}, {J.W. K\l{}os}, {S. Mamica},
  {M. Krawczyk}, Magnonic crystals - prospective structures for shaping spin
  waves in nanoscale, Low Temperature Physics 41 (2015) 10.

\end{thebibliography}

\end{document}